\documentclass{article}

\usepackage{microtype}
\usepackage{graphicx}
\usepackage{subfigure}
\usepackage{amsmath}
\usepackage{amsfonts}
\usepackage{booktabs,multirow,graphicx,xcolor,colortbl}
\usepackage{xcolor}
\usepackage{xurl}
\definecolor{oursgreen}{RGB}{230,250,230}
\definecolor{bestgreen}{RGB}{0,140,70}
\newcommand{\best}[1]{\textcolor{bestgreen}{\textbf{#1}}}
\DeclareRobustCommand{\flowstep}[2]{%
    \hyperref[#2]{\textbf{(#1)}}%
}
\newcommand{\perfbest}[1]{\textcolor{green!45!black}{\textbf{#1}}}

\usepackage{hyperref}

\usepackage[accepted]{mlsys2026}
\usepackage{etoolbox}
\makeatletter
\renewcommand{\Notice@String}{Preprint.}
\patchcmd{\printAffiliationsAndNotice}
  {{\bf AUTHORERR: Missing \textbackslash{}mlsyscorrespondingauthor.}}
  {}{}{\PackageError{coffee}{Could not configure preprint affiliations}{}}
\makeatother
\hypersetup{
  pdftitle={Elastic Precision: Software--Hardware Co-Design of a Mixed-Precision Architecture for Long-Context LLM Inference},
  pdfauthor={Boyang Chen, Xiao Wu, Haoran Wu, Yiren Zhao},
  pdfsubject={Preprint}
}

\mlsystitlerunning{Elastic Precision for Long-Context LLM Inference}

\begin{document}
\raggedbottom
\twocolumn[
\mlsystitle{Elastic Precision: Software--Hardware Co-Design of a Mixed-Precision Architecture for Long-Context LLM Inference}

\begin{mlsysauthorlist}
\mlsysauthor{Boyang Chen}{imperial}
\mlsysauthor{Xiao Wu}{imperial}
\mlsysauthor{Haoran Wu}{cambridge}
\mlsysauthor{Yiren Zhao}{imperial}
\end{mlsysauthorlist}

\mlsysaffiliation{imperial}{Imperial College London, London, United Kingdom}
\mlsysaffiliation{cambridge}{University of Cambridge, Cambridge, United Kingdom}

\mlsyskeywords{Long-context LLM inference, Mixed-precision quantization, Runtime precision allocation, Hardware accelerators, Software--hardware co-design, KV-cache optimization}

\vskip 0.3in

\begin{abstract}
Low-bit quantization reduces the data movement and
computation costs of large language model (LLM) inference.
However, uniform widths can waste precision and degrade
model quality because computations differ in quantization
sensitivity.
Existing mixed-precision methods alleviate this mismatch
by assigning widths to channels, tokens, or quantization
groups, with some adapting widths to runtime inputs.
However, when selecting one input's width, these methods
do not fully account for how the other input of the
current GEMM tile affects output error.
Combining both inputs of the current GEMM tile to
estimate output errors and allocate both input widths
can therefore better match precision to the current
computation, but introduces additional estimation and
selection overheads on conventional GPUs and NPUs.
We therefore propose Coffee, a runtime precision
allocation architecture for long-context LLM inference.
Under a fixed average bit-width budget, Coffee combines
information from both input matrices of the current
GEMM tile to estimate output errors at candidate widths.
These estimates guide online allocation of both
input widths across subarrays, directing more bits
toward computations with greater estimated error
reductions.
Its mixed-precision hardware directly supports
the selected widths in data access and computation,
while overlapping error estimation and width selection
with transfers and computation to limit additional stalls.
The same read path supports attention-guided KV reads
during decoding, reducing traffic and easing
bandwidth bottlenecks.
Coffee achieves lower perplexity and higher average
zero-shot accuracy than the evaluated quantization
baselines, and higher long-context retrieval recall
than PLENA.
For long-context decoding, system modeling projects
up to $2.55\times$ the throughput and $2.61\times$
the energy efficiency of the reconfigured PLENA baseline,
and up to $4.96\times$ and $12.56\times$, respectively,
compared with the B200 GPU baseline.
\end{abstract}

]

\printAffiliationsAndNotice{}

\section{Introduction}
\label{sec:introduction}

Large language model (LLM) inference relies heavily
on matrix multiplications in linear projections
and attention.
Low-bit quantization reduces their data movement
and computation costs, but the resulting errors
can degrade model quality.
For example, on a single H100 at batch size one,
NVIDIA reports that AWQ INT4 weight-only quantization
raises Llama-3-8B throughput to $1.56\times$ that of FP16
while reducing MMLU accuracy from 65.4\% to
62.9\%~\cite{nvidia_ptq2024}.

Prior work mitigates quantization-induced quality loss
through calibration-based error optimization~\cite{gptq},
data transformations~\cite{quarot,flatquant}, and
adaptive numerical representations~\cite{ant,mant}.
Mixed-precision methods complement these techniques
by assigning more bits to sensitive computations.
APTQ~\cite{aptq} and CMPQ~\cite{cmpq} allocate widths
offline at layer or channel granularity using
sensitivity or importance estimates.
Other methods adapt precision during inference:
FGMP~\cite{fgmp} selects activation widths online,
while OPAL~\cite{opal} preserves runtime activation
outliers at higher precision.
Both retain offline-assigned weight precision.
MoBiQuant~\cite{mobiquant} instead selects weight
precision for individual tokens while keeping
activation widths fixed.

\begin{figure}[t]
    \centering

    \begin{minipage}[t]{0.48\linewidth}
        \vspace{0pt}
        \centering
        \includegraphics[width=\linewidth]
            {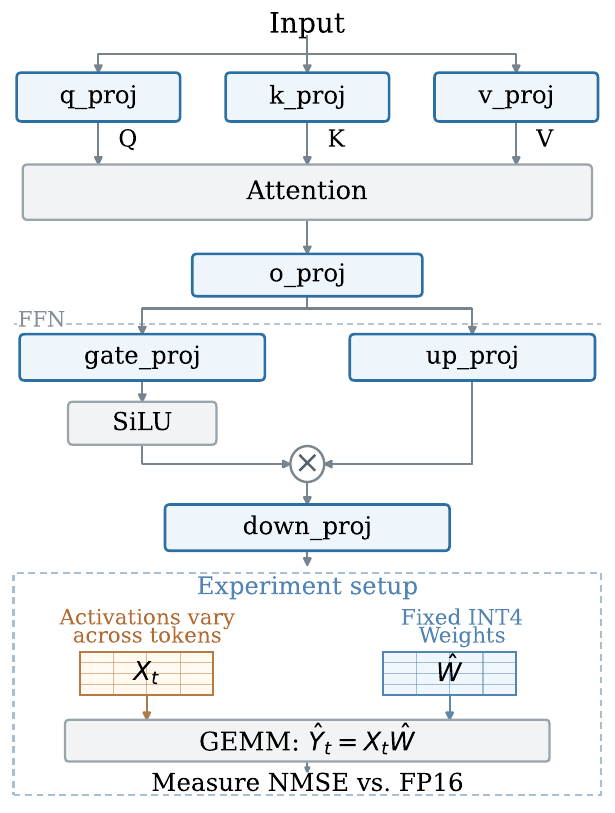}
        \par\smallskip
        {\small\textbf{(a) Experimental setup.}\par}
    \end{minipage}
    \hfill
    \begin{minipage}[t]{0.48\linewidth}
        \vspace{0pt}
        \centering
        \includegraphics[width=\linewidth]
            {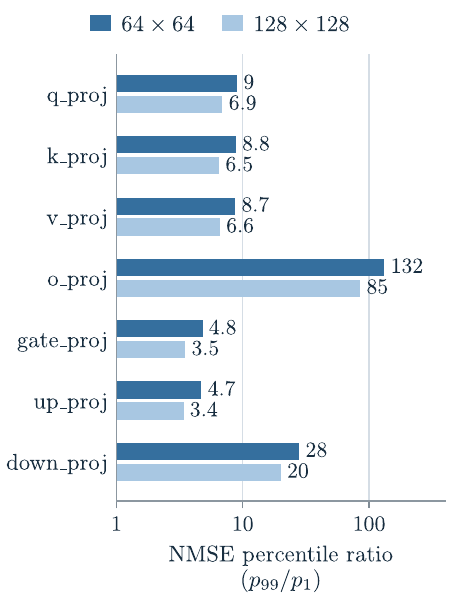}
        \par\smallskip
        {\small\textbf{(b) Results.}\par}
    \end{minipage}

    \caption{
    Input-dependent output error with fixed INT4 weights.
    \textbf{(a) Experimental setup.}
    The diagram locates the seven linear operators in
    Llama-3.1-8B.
    Their GEMMs are mapped to $64\times64$ and
    $128\times128$ matrix-multiplication units.
    For each unit, we fix the INT4 weight representation
    and vary FP16 activations across tokens.
    \textbf{(b) Results.}
    For each operator and unit size, bars report the
    99th-to-1st percentile ratio ($p_{99}/p_1$) of
    output normalized mean squared error (NMSE)
    across tokens.
    NMSE is $\|\widehat{Y}-Y\|_F^2/\|Y\|_F^2$,
    where $Y$ and $\widehat{Y}$ are computed using
    the same FP16 activations with FP16 and INT4
    weights, respectively.
    These results show that even when one input's
    quantized representation remains fixed, the output
    error caused by its quantization still depends
    on the other input.
    }
    \label{fig:operand_error_variation}
\end{figure}

These runtime methods adapt one input's precision
while leaving the other's unchanged.
However, the output error from quantizing either
input depends on both inputs of the current GEMM tile.
For example, Figure~\ref{fig:operand_error_variation}
shows that the same INT4 weights produce different
output errors when multiplied by different token
activations.
Likewise, the output error from activation
quantization depends on the corresponding weights.
Selecting either input's width should therefore
account for the current input on the other side;
otherwise, additional bits may not be allocated
where they would most effectively reduce output error.
Coffee combines both current inputs to estimate
output errors at candidate widths and uses these
estimates to select each subarray's two input widths.
Within the assigned budgets, it allocates additional
bits where they yield greater estimated reductions
in output error.

However, implementing such allocation on conventional
GPUs and NPUs requires retrieving or estimating
both inputs' magnitudes and quantization errors
at candidate widths, then using this information
to estimate output errors and select widths for
each tile.
These steps add instructions, data accesses,
and computation.
Separate GPU kernels also incur launch overhead and
intermediate data transfers.
GPU matrix units further support only a limited set of
predefined formats~\cite{nvidia_ptx}, making it difficult
to translate one-bit width adjustments and different
input-width combinations into execution benefits.
These overheads and format constraints motivate
hardware that makes low-overhead decisions and supports
the selected widths in data access and computation.

We therefore propose \textbf{Coffee}, a hardware architecture for runtime
precision allocation in long-context LLM inference.
Its compute core follows PLENA's flattened systolic-array
organization~\cite{plena}, distributing matrix multiplication
along the reduction dimension across subarrays with
independently configurable widths for each subarray's
two inputs.
Dedicated circuitry overlaps output-error estimation
and width selection with input transfers and preceding
computations to limit additional stalls.
Subarray gating and selective reads from nested encodings
translate lower widths into reduced compute energy
and data movement.

Coffee combines offline budget allocation with online
width selection.
Offline calibration estimates sensitivity differences
across computations and assigns more budget to more
sensitive computations under a unified average
bit-width constraint.
Since these estimates do not fully capture local
requirements that vary with current inputs, the runtime
pipeline combines information from both current input
matrices to estimate output errors at candidate widths.
These estimates guide the selection of each subarray's
two input widths within the assigned tile input-side
budgets, allocating additional bits where they yield
greater estimated reductions in output error.
The same mixed-precision read path also supports
attention-guided KV reads during long-context decoding,
building on prior importance-aware KV
quantization~\cite{mikv,zipcache} to further reduce
KV traffic.

Our main contributions are as follows:

\begin{itemize}
    \item \textbf{Dynamic precision allocation guided
    by both current inputs.}
    We use both current inputs to estimate output errors
    and guide online selection of each subarray's two
    input widths within offline-assigned budgets,
    prioritizing precision increases with greater
    estimated error reductions.

    \item \textbf{Efficient hardware for runtime
    mixed-precision execution.}
    We integrate online decision logic, independent
    input-width control, subarray gating, and selective
    reads from nested encodings.
    The shared hardware supports dynamic precision allocation
    and attention-guided KV reads, translating low precision
    into lower compute energy and reduced data movement.

    \item \textbf{Improved model quality and system efficiency.}
    Our method achieves lower perplexity and higher average
    zero-shot accuracy than the evaluated quantization
    baselines, and higher long-context retrieval recall
    than PLENA.
    For long-context decoding, system modeling projects
    up to $2.55\times$ the throughput and $2.61\times$
    the energy efficiency of the reconfigured PLENA baseline,
    and up to $4.96\times$ and $12.56\times$, respectively,
    compared with the B200 GPU baseline.
\end{itemize}

\section{Related Work}
\label{sec:related_work}

\subsection{LLM Quantization}

GPTQ~\cite{gptq} uses second-order approximations
for offline weight quantization.
QuaRot~\cite{quarot} and FlatQuant~\cite{flatquant}
transform data to reduce quantization error, while
ANT~\cite{ant} and M-ANT~\cite{mant} adapt numerical formats.
APTQ~\cite{aptq} and CMPQ~\cite{cmpq} allocate widths
using offline sensitivity or importance estimates.
FGMP~\cite{fgmp} selects activation widths online,
while OPAL~\cite{opal} preserves runtime activation
outliers at higher precision.
Both retain offline-assigned weight precision.
MoBiQuant~\cite{mobiquant} dynamically selects weight
precision for each token while keeping activation
widths fixed.
For KV caches, KIVI~\cite{kivi} quantizes keys per
channel and values per token, while MiKV~\cite{mikv}
and ZipCache~\cite{zipcache} assign precision by
token importance.
Coffee uses both current input matrices to
select both local input widths online within
assigned budgets.

\subsection{Flattened Systolic Array}

PLENA~\cite{plena} distributes matrix multiplication
across subarrays along the reduction dimension
and combines their partial sums through an adder tree.
Coffee adopts this organization and adds online
precision selection, independent input-width control,
gating, and selective reads from nested encodings
to support runtime mixed-precision execution.

\section{Motivation}
\subsection{Error Proxy Derivation}
\label{sec:error_derivation}

We construct an output-error proxy that accounts
for both current input matrices.
For subarray $i$ in the flattened systolic array,
let $A_i$ and $B_i$ denote its inputs, with quantization
errors $\Delta A_i$ and $\Delta B_i$.
The local output $C_i=A_iB_i$ has quantization error
\begin{equation}
\begin{aligned}
\Delta C_i
&=(A_i+\Delta A_i)(B_i+\Delta B_i)-A_iB_i\\
&=A_i\Delta B_i+\Delta A_iB_i+\Delta A_i\Delta B_i.
\end{aligned}
\label{eq:matrix_output_error}
\end{equation}
The first two terms propagate each input's quantization
error through the other input.
Thus, output error can vary with the paired input
even when one input and its quantization error remain fixed.
The last term captures the interaction between
the two quantization errors.

Let $P_i^A$ and $P_i^B$ denote the inputs' mean-square
magnitudes, and $\varepsilon_i^A(w_A)$ and
$\varepsilon_i^B(w_B)$ their quantization MSEs
at candidate widths $w_A$ and $w_B$.
Assuming zero-mean quantization noise independent
of the inputs and one another, we use these input
statistics (mean-square magnitudes and quantization
MSEs) to construct the following output-error proxy:
\begin{equation}
\begin{aligned}
R_i(w_A,w_B)
={}&P_i^B\varepsilon_i^A(w_A)
+P_i^A\varepsilon_i^B(w_B)\\
&+\varepsilon_i^A(w_A)\varepsilon_i^B(w_B).
\end{aligned}
\label{eq:full_error_proxy}
\end{equation}

When quantization MSEs are small relative to the
corresponding signal mean-square magnitudes,
the product term is negligible, giving the
separable approximation
\begin{equation}
\widehat R_i(w_A,w_B)
=
P_i^B\varepsilon_i^A(w_A)
+
P_i^A\varepsilon_i^B(w_B).
\label{eq:separable_error_proxy}
\end{equation}
The two input contributions are therefore
\begin{align}
r_i^A(w_A)&=P_i^B\varepsilon_i^A(w_A),
\label{eq:operand_a_error}\\
r_i^B(w_B)&=P_i^A\varepsilon_i^B(w_B).
\label{eq:operand_b_error}
\end{align}
With $P_i^A$ and $P_i^B$ fixed for the current input pair,
each error contribution depends only on its corresponding input width.

\begin{figure}[t]
    \centering
    \includegraphics[width=\linewidth]
        {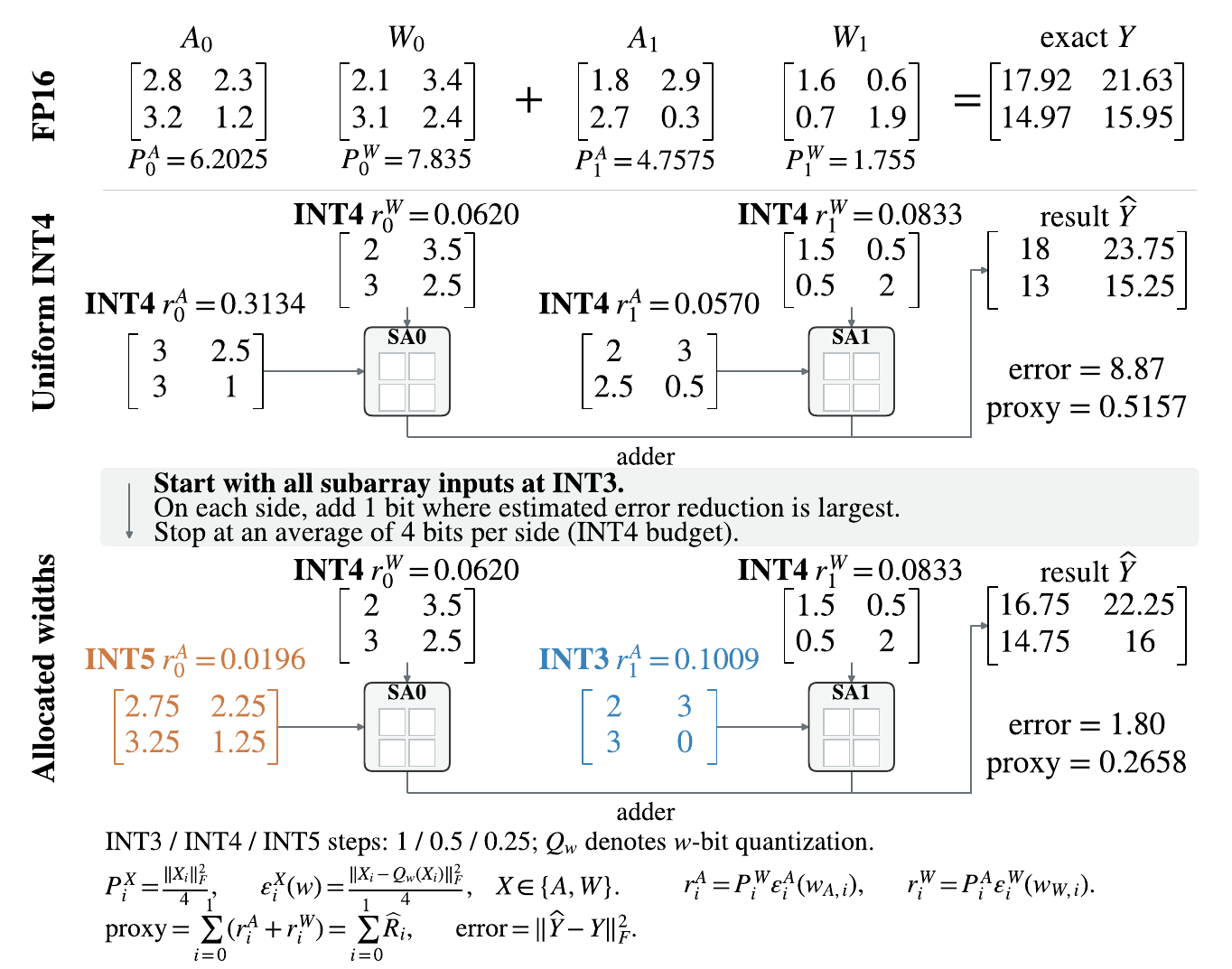}
    \caption{
Proxy-guided precision redistribution on a $2\times4$
flattened systolic array comprising two $2\times2$
subarrays, computing $Y=A_0W_0+A_1W_1$.
Statistics from both current input matrices guide
allocation through Eqs.~\eqref{eq:operand_a_error}
and~\eqref{eq:operand_b_error}.
Starting from uniform INT3, each input side repeatedly
applies the eligible one-bit upgrade with the largest
estimated error reduction until its average width
reaches four bits.
Relative to uniform INT4, one bit of activation precision
moves from SA1 to SA0, yielding INT5 and INT3
activations, respectively, while both weight inputs
remain INT4.
The total separable proxy decreases from 0.5157
to 0.2658 (48.5\%), and the squared output error
from 8.87 to 1.80 (79.7\%).
    }
    \label{fig:nonuniform_precision}
\end{figure}
\begin{figure}[t]
    \centering
    \includegraphics[width=\linewidth]
        {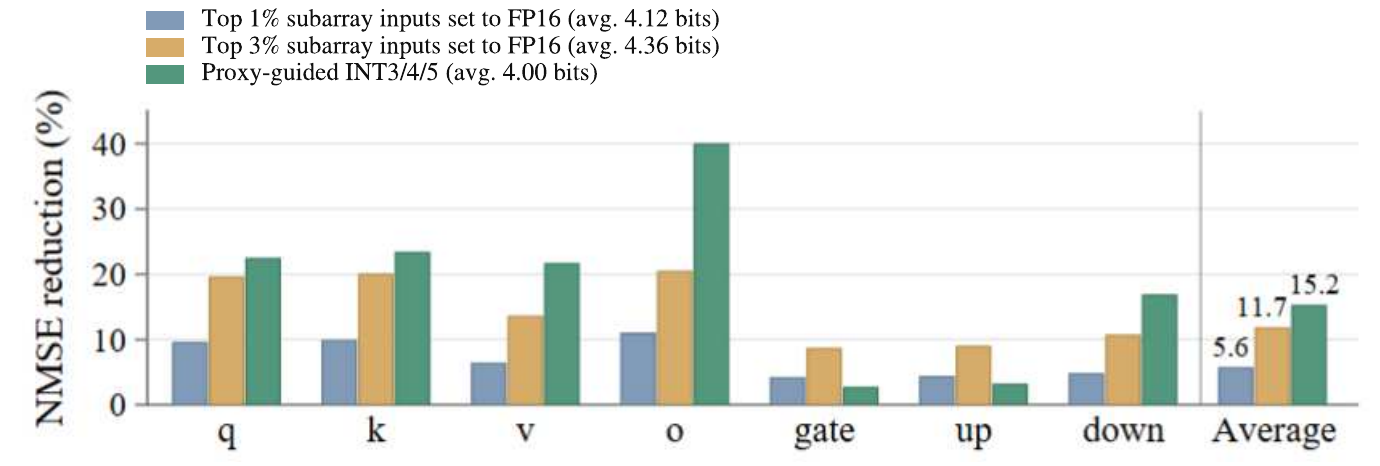}
    \caption{
Output NMSE reduction relative to uniform INT4
on Llama-3-8B.
We map the seven linear operators across all 32 layers
to a $16\times2048$ flattened systolic array.
Each GEMM is independently quantized and evaluated
under each policy using the same inputs captured
from an FP16 run.
Proxy-guided INT3/4/5 allocation follows
Figure~\ref{fig:nonuniform_precision}
under an average four-bit budget.
Outlier policies rank subarray inputs by maximum absolute
value separately for each input side, assigning FP16
to the top 1\% or 3\% and INT4 to the rest.
Average reports the relative reduction in MAC-weighted
NMSE: 15.2\% for proxy-guided allocation, versus
5.6\% and 11.7\% for the two outlier policies,
whose average widths are 4.12 and 4.36 bits, respectively.
    }
    \label{fig:nonuniform_precision_reduction}
\end{figure}
\subsection{Precision Allocation Using the Error Proxy}
\label{sec:nonuniform_precision}

To evaluate the proxy's effectiveness, we first use
the separable error proxy $\widehat R_i$ to redistribute
precision on a $2\times4$ flattened systolic array
comprising two $2\times2$ subarrays (SA0 and SA1).
The allocation uses the current input matrices while
maintaining an average of four bits per input side.
Figure~\ref{fig:nonuniform_precision} shows that
this allocation reduces the total proxy from
0.5157 to 0.2658 and the squared output error
from 8.87 to 1.80, a 79.7\% reduction relative
to uniform INT4.

We further evaluate this allocation method on
Llama-3-8B, comparing proxy-guided INT3/4/5 allocation
with two outlier-based FP16-retention policies
(Figure~\ref{fig:nonuniform_precision_reduction}).
Using statistics from both current input matrices,
our allocation reduces average output NMSE by 15.2\%
at an average width of four bits, outperforming
both outlier policies despite their higher average widths.

\section{Precision Allocation Framework}
\label{sec:precision_allocation}

\begin{figure*}[t]
    \centering
    \includegraphics[width=0.7\linewidth]
        {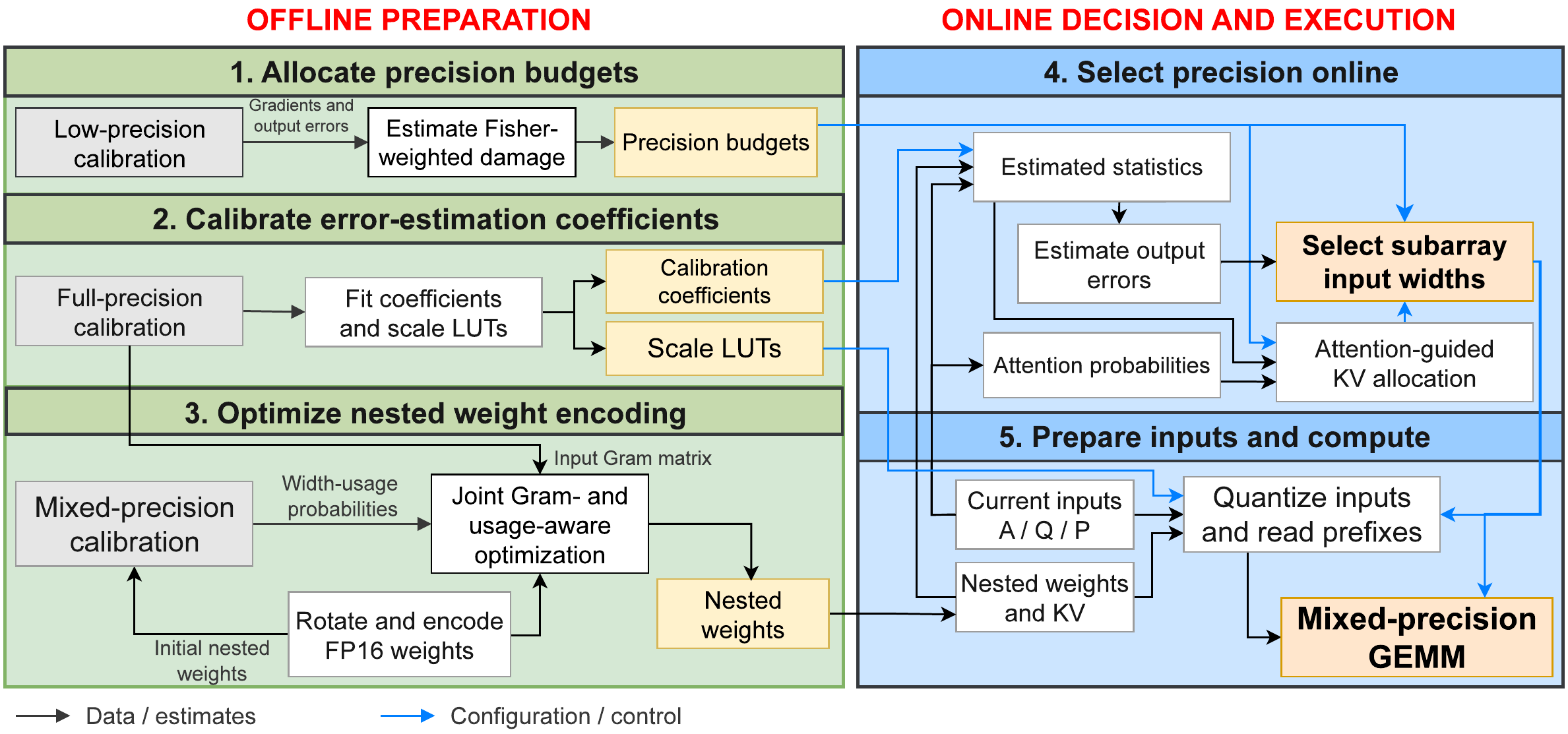}
\caption[Offline preparation and online execution.]{
Offline preparation and online execution.
Offline,
\textbf{(1)} low-precision calibration provides gradients
and output errors for \hyperref[flow:budget]{Fisher-guided
budget allocation};
\textbf{(2)} full-precision calibration fits
\hyperref[flow:calibration]{error-estimation coefficients}
and \hyperref[flow:lut]{scale LUTs};
and \textbf{(3)} \hyperref[flow:encoding]{nested weight encoding}
jointly optimizes shared exponents and codes using
input Gram matrices and width-usage probabilities.
Online,
\textbf{(4)} both inputs' estimated mean-square magnitudes
and reconstruction MSEs guide
\hyperref[sec:online_selection]{subarray width selection}
within the assigned budgets, with attention probabilities
additionally guiding \hyperref[sec:kv_read_selection]{KV allocation};
and \textbf{(5)} selected widths control input quantization,
nested-prefix reads, and \hyperref[sec:mixed_precision_arrays]{mixed-precision GEMM}.
Black and blue arrows denote data/estimates and
configuration/control, respectively.
}
    \label{fig:precision_allocation_flow}
\end{figure*}

Coffee assigns precision budgets offline under a unified
average bit-width constraint.
At runtime, it combines information from both current
input matrices to select subarray input widths within
these budgets.
Attention information further guides KV read precision
during decode.
Figure~\ref{fig:precision_allocation_flow} summarizes
the framework.

\subsection{Online Error Estimation and Bit-Width Selection}
\label{sec:online_selection}

Evaluating the output-error contributions in
Eqs.~\eqref{eq:operand_a_error}
and~\eqref{eq:operand_b_error} at runtime requires
input mean-square magnitudes and reconstruction
MSEs at candidate widths.
To avoid costly elementwise squaring, accumulation,
and repeated quantization, Coffee estimates these
statistics at low cost.

Each subarray input contains multiple MXINT quantization
groups, each sharing an exponent $e$ that may differ
across groups.
For a group of $n$ values $x_j$, let
$a_{\max}=\max_j|x_j|$.
We estimate its mean-square magnitude as
\begin{equation}
P=\frac{1}{n}\sum_{j=1}^{n}x_j^2
\approx\kappa_{\mathrm{op}}a_{\max}^2
=\widehat P.
\label{eq:online_magnitude_estimate}
\end{equation}
For each operator type $\mathrm{op}$,
$\kappa_{\mathrm{op}}$ is the mean of $P/a_{\max}^2$
over calibration groups.

Let $2^e$ denote the group's quantization step at
the maximum supported width $w_{\max}$.
With $e$ fixed, each one-bit reduction doubles
the reconstruction spacing, giving at width $w$
\begin{equation}
\Delta_w=2^{e+w_{\max}-w}.
\label{eq:online_reconstruction_spacing}
\end{equation}
Let $\varepsilon(w)$ denote the group's reconstruction
MSE at width $w$.
Approximating rounding noise as uniform over
$[-\Delta_w/2,\Delta_w/2]$ and neglecting saturation
gives $\varepsilon(w)\approx\Delta_w^2/12$
~\cite{gray1998quantization}.
Since reconstructing the group as zero gives MSE $P$,
we empirically cap this analytical estimate at
$\widehat P$ before applying a calibration coefficient:
\begin{equation}
\widehat\varepsilon(w)
=
c_{\mathrm{op},w}
\min\left(
\frac{\Delta_w^2}{12},
\,\widehat P
\right).
\label{eq:online_mse_estimate}
\end{equation}
\phantomsection\label{flow:calibration}
We fit $c_{\mathrm{op},w}$ to measured reconstruction
MSEs for each operator type and candidate width.

\phantomsection\label{flow:estimation}
For subarray $i$ and input side $s\in\{A,B\}$,
element-count-weighted averages of the group estimates
give $\widehat P_i^s$ and
$\widehat\varepsilon_i^s(w)$.
Weight statistics are computed offline; activation
and KV statistics are estimated before memory writes.
These estimates, including MSEs at all candidate widths,
are stored alongside the data and combined to estimate
each side's output-error contribution:
\begin{equation}
\widehat r_i^A(w)
=
\widehat P_i^B\,\widehat\varepsilon_i^A(w),
\qquad
\widehat r_i^B(w)
=
\widehat P_i^A\,\widehat\varepsilon_i^B(w).
\label{eq:online_input_error}
\end{equation}

For a GEMM tile mapped to $M$ subarrays with equal-sized
inputs on each side, the average-width budget
$\bar w_s$ gives a total width budget
$\mathcal B_s=M\bar w_s$ for side $s$.
Budgets are assigned offline
(Section~\ref{sec:offline_budget_allocation});
decode K/V is handled separately in
Section~\ref{sec:kv_read_selection}.
Let $w_i^s$ be the selected input width from the supported
integer set $\mathcal W=\{w_{\min},\ldots,w_{\max}\}$.
For each input side $s$, we select widths under
its assigned budget with the objective
\begin{equation}
\min_{w_i^s\in\mathcal W}
\sum_{i=1}^{M}\widehat r_i^s(w_i^s)
\quad\text{s.t.}\quad
\sum_{i=1}^{M}w_i^s\leq\mathcal B_s.
\label{eq:online_width_allocation}
\end{equation}
The marginal gain from a one-bit upgrade is
\begin{equation}
g_{i,w\rightarrow w+1}^{s}
=
\widehat r_i^s(w)-\widehat r_i^s(w+1),
\qquad w<w_{\max}.
\label{eq:online_upgrade_gain}
\end{equation}
\phantomsection\label{flow:selection}
We use a greedy heuristic, starting all subarray
inputs at $w_{\min}$ and repeatedly applying the
largest-gain one-bit upgrade available from the
current widths until the remaining budget is insufficient
or all inputs reach $w_{\max}$.

\subsection{Attention-Guided KV Precision Allocation}
\label{sec:kv_read_selection}

During decode, numerical distributions~\cite{kivi}
alone do not capture historical tokens' query-dependent
contributions to the attention output.
Coffee combines attention information~\cite{h2o}
with reconstruction-error estimates to guide KV reads,
using current Q inputs for K allocation and current
P inputs for V allocation.
Read widths are assigned at subarray granularity.
Offline planning sets per-layer proportions for K/V
read widths within $[w_{\min},w_{\max}]$ and V skipping
(Section~\ref{sec:offline_budget_allocation}).

Before QK computation, current attention probabilities
are unavailable, so K allocation uses the preceding
query's buffered softmax probabilities $\hat p_{h,t}$.
Let $\mathcal T_i$ and $\beta_i$ denote the token indices
and channel group of subarray $i$'s K input.
For layer $\ell$ and KV head $k$, its estimated damage
at read width $w$ is
\begin{equation}
\widehat d_i^K(w)
=
G_{\ell,k}^K\alpha^2
\sum_{h\in\mathcal H_k}
\widehat U_{h,\beta_i}
\sum_{t\in\mathcal T_i}
\hat p_{h,t}^{\,2}
\widehat\varepsilon^K_{t,\beta_i}(w),
\label{eq:decode_k_damage}
\end{equation}
where $G_{\ell,k}^{K/V}$ are offline-calibrated
sensitivity coefficients, $\alpha$ is the attention
scaling factor, and $\mathcal H_k$ contains the query
heads sharing KV head $k$.
The current query-group energy is estimated as
$\widehat U_{h,\beta_i}
=n_{\beta_i}\widehat P^Q_{h,\beta_i}$,
where $n_{\beta_i}$ is the group's channel count
and $\widehat P^Q_{h,\beta_i}$ follows
Eq.~\eqref{eq:online_magnitude_estimate}.
This energy weights K errors propagated through Q
and is reused across tokens and candidate widths.
The per-element K reconstruction MSE
$\widehat\varepsilon^K_{t,\beta_i}(w)$ is estimated
using Eq.~\eqref{eq:online_mse_estimate}
before K writes and stored with the data.

K subarrays containing recent tokens or outlier data
are protected at prescribed widths; their counts
are deducted from the per-layer quotas.
Remaining quotas are filled from high to low widths,
selecting unassigned subarrays with the largest
$\widehat d_i^K(w)$ at each width $w$.
All K subarrays are read.
The last prefill query's attention probabilities
initialize $\hat p$ for the first decode step.
With K widths fixed, Q widths are selected using
Section~\ref{sec:online_selection}.

After softmax, V allocation uses current attention
probabilities $P_{h,t}$.
Let $\mathcal T_i$ denote the token indices in
subarray $i$'s V input.
The estimated read and skip damages are
\begin{equation}
\begin{aligned}
\widehat d_i^V(w)
&=
G_{\ell,k}^V
\left(
\sum_{h\in\mathcal H_k}
\sum_{t\in\mathcal T_i}P_{h,t}^{\,2}
\right)
\widehat\varepsilon_i^V(w),\\
\widehat d_i^{V,\mathrm{skip}}
&=
G_{\ell,k}^V
\left(
\sum_{h\in\mathcal H_k}
\sum_{t\in\mathcal T_i}P_{h,t}^{\,2}
\right)
\widehat P_i^V.
\end{aligned}
\label{eq:decode_v_damage}
\end{equation}
Here, $\widehat\varepsilon_i^V(w)$ and $\widehat P_i^V$
are the estimated reconstruction MSE and mean-square
magnitude, obtained by averaging group estimates from
Eqs.~\eqref{eq:online_mse_estimate}
and~\eqref{eq:online_magnitude_estimate}, respectively,
before V writes and stored with the data.
The squared-probability sum is computed once per step
and reused across widths and skip decisions.

V subarrays containing recent tokens or exhibiting
high V energy are protected at prescribed widths,
with their counts deducted from the per-layer quotas.
Among unprotected subarrays, those with the smallest
$\widehat d_i^{V,\mathrm{skip}}$ are skipped to meet
the prescribed proportion.
Retained subarrays fill the remaining width quotas
as in K allocation, using $\widehat d_i^V(w)$ for ranking.
With V widths fixed, P widths are selected for retained
subarrays using Section~\ref{sec:online_selection}.
Skipping omits V reads and the corresponding P/V-port
operations without renormalizing attention probabilities.

\subsection{Offline Precision-Budget Allocation}
\label{sec:offline_budget_allocation}
\phantomsection\label{flow:budget}

We calibrate prefill and decode separately at each
context length.
Precision cost $C$ sums bit widths weighted by
subarray input-port use counts.

Following BRECQ~\cite{brecq}, we estimate quantization
damage with a diagonal empirical-Fisher approximation:
\begin{equation}
D_i^s(w)
=
\sum_{t,j}
g_{i,t,j}^{2}
\bigl(E_i^s(w)_{t,j}\bigr)^2,
\label{eq:offline_damage}
\end{equation}
where $i$ indexes GEMM tiles, $s\in\{A,B\}$ denotes
the input side, and $t,j$ index operator-output entries.
Gradients $g_{i,t,j}$ are computed from the calibration
loss at the corresponding operator outputs in a
low-precision model.
$E_i^s(w)$ is the output error with side $s$ at width
$w$ and the other side at reference precision;
weight and KV errors use actual nested-prefix
reconstructions.
To compare computations on a common mean negative
log-likelihood (NLL) scale, we normalize upgrade gains
$\tfrac12[D_i^s(w)-D_i^s(w+1)]$ per valid prediction
token and fit them to measured mean NLL changes.
Total damage $\widehat D$ sums the calibrated contributions.

For prefill, the objective is
\begin{equation}
\min_{\pi}\widehat D(\pi)
\quad\text{s.t.}\quad
C(\pi)\leq\bar wN_0,
\label{eq:offline_budget_allocation}
\end{equation}
where $\pi$ is the allocation, $\bar w$ the target
average width, and $N_0$ the fixed valid input-port
use count.
The planner starts from minimum widths and greedily
prioritizes damage reduction per unit precision cost.

For decode, the planner also considers subarray-level
K/V read widths and V skipping, with their
attention-dependent damages calibrated to the same
NLL scale.
Skipping changes the executed input-port use count
$N_{\mathrm{exec}}(\pi)$, so the budget becomes
\begin{equation}
C(\pi)\leq\bar wN_{\mathrm{exec}}(\pi).
\label{eq:decode_executed_budget}
\end{equation}
A Lagrangian relaxation guides the joint discrete search:
\begin{equation}
\begin{aligned}
J_\lambda(\pi)
={}&\widehat D(\pi)\\
&+\lambda\left[
C(\pi)-\bar wN_{\mathrm{exec}}(\pi)
\right],
\qquad \lambda\geq0.
\end{aligned}
\label{eq:decode_allocation_score}
\end{equation}
For each $\lambda$, shared across all operators,
the planner jointly searches linear-operator, Q/P,
and K/V configurations using $J_\lambda$.
We adjust $\lambda$ by bisection on calibration inputs
to find budget-feasible configurations near the target
executed-port average width, then verify
Eq.~\eqref{eq:decode_executed_budget}.
Skipped P/V operations count toward neither $C$
nor $N_{\mathrm{exec}}$.

The planner returns tile input-side budgets for online
selection in linear operators and Q/P, and aggregates
K/V subarray assignments across calibration inputs
into per-layer proportions for online allocation.

\section{Hardware Support for Mixed-Precision Execution}

Dedicated circuitry overlaps runtime precision selection
with data transfers and preceding computations
to limit additional stalls.
Selective nested reads and subarray gating translate
the selected widths into reduced data movement
and compute energy.

\subsection{Architecture and Execution Flow}
\label{sec:architecture_execution}

\begin{figure}[t]
    \centering
    \includegraphics[width=\linewidth]
        {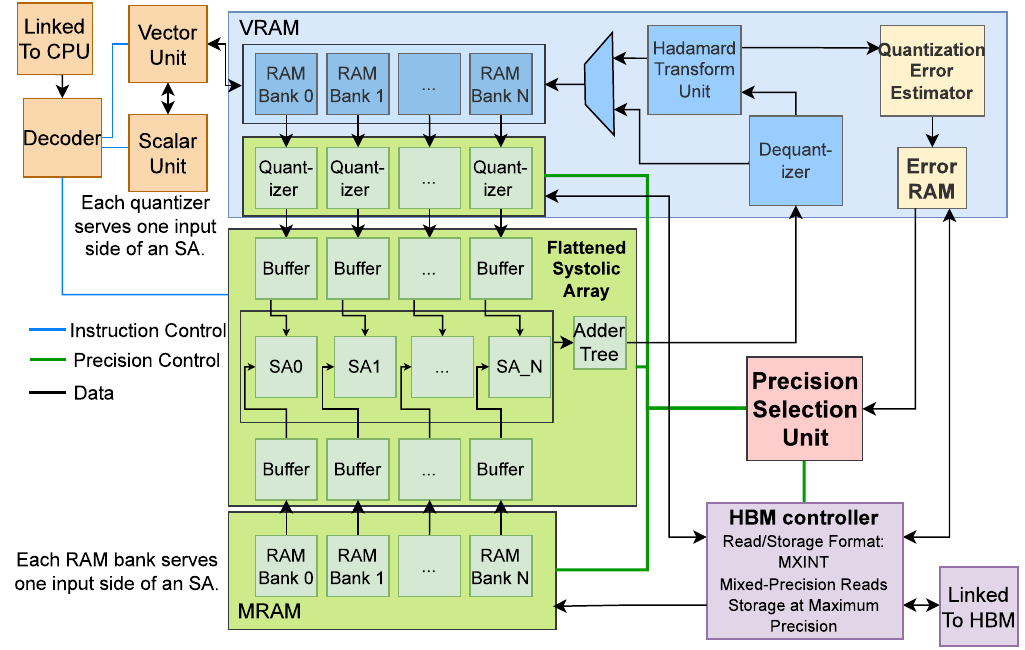}
    \caption{A Coffee compute core and its memory interfaces;
    see Section~\ref{sec:architecture_execution} for details.}
    \label{fig:architecture}
\end{figure}

Figure~\ref{fig:architecture} shows a Coffee compute core
and its memory interfaces.
The core adopts PLENA's flattened systolic-array
organization~\cite{plena}, with separate data and
precision-control paths for each subarray's two inputs.
The Precision Selection Unit (PSU) determines their
widths online.
An instruction decoder coordinates execution, while
vector and scalar units handle non-matrix operations.

The array receives inputs through banked Vector SRAM
(VRAM) and Matrix SRAM (MRAM), with each data bank
serving one input side of a corresponding subarray.
VRAM stores FP16 activations, Q/P, and intermediate
results, and quantizes outgoing inputs through
a configurable quantizer per bank.
MRAM buffers nested weight and KV codes from HBM
and supplies the selected prefixes.
A dedicated VRAM bank, Error RAM, buffers the estimates
used by the PSU.
Weight codes, shared exponents, and error and magnitude
statistics are prepared offline.

Preparation for each GEMM begins when preceding
computations produce its inputs.
Activations, Q/P, and newly generated K/V return
to VRAM in FP16.
Along the writeback path, the Hadamard Transform Unit
(HTU) performs the required rotations for linear GEMMs
and QK, while P and V bypass rotation.
The estimator then prepares mean-square magnitudes
and reconstruction MSEs at candidate widths
for the resulting inputs
(Section~\ref{sec:online_selection}).
The vector unit also computes and buffers attention
probabilities for subsequent KV precision allocation.

During prefill, inputs and their associated estimates
are written back to HBM when on-chip capacity
is insufficient.
During decode, newly generated K/V and their estimates
are written to HBM, while activations and their
estimates remain on chip.
KV data use nested encodings for storage.
Reconstruction-MSE estimates use INT6 throughout,
including on-chip buffering and HBM writeback.

\phantomsection\label{flow:execution}
Before the next GEMM, the required estimates are
gathered in Error RAM.
For prefill inputs stored in HBM, these estimates
are prefetched ahead of execution.
Decode linear GEMMs use estimates already on chip,
whereas KV allocation retrieves error and magnitude
estimates from HBM.
The PSU combines information from both inputs to
select subarray widths within the assigned budgets,
using attention probabilities additionally for
decode KV allocation
(Sections~\ref{sec:online_selection}
and~\ref{sec:kv_read_selection}).
This selection precedes precision-dependent data reads.

Once the widths are determined, VRAM quantizers
quantize the current inputs, and the HBM controller
requests the corresponding weight/KV bit planes.
MRAM forwards the selected prefixes and shared
exponents to the array.
Input buffers hold both inputs until the data and
width configurations are ready, after which the
subarrays execute with the corresponding gating.
Their partial results are scale-aligned, reduced
through an adder tree, and dequantized to FP16
for VRAM writeback.

\subsection{Rotation and Adaptive Online Quantization}

\phantomsection\label{flow:rotation}
For GEMMs other than $PV$, Coffee applies matched Hadamard
rotations along each subarray's reduction dimension,
preserving the unquantized product while spreading outliers.
The rotation granularity matches the shared-exponent
quantization groups.
Weights are rotated and encoded offline; runtime inputs
requiring rotation pass through the HTU before VRAM writes.
Rotated data are reused across computations and width changes.

\phantomsection\label{flow:lut}
Adaptive quantization selects each quantization group's
shared exponent using an offline-calibrated LUT.
Threshold comparisons bin features derived from the
two largest absolute values and the group magnitude
relative to a calibrated reference.
The LUT returns an exponent adjustment calibrated from
reconstruction errors under candidate adjustments,
balancing rounding and clipping without runtime scale
searches or repeated quantization.
For A/Q/P, scale selection targets the selected width.
For KV, calibration weights reconstruction errors across
read widths by their expected usage frequencies.
Each KV group's exponent is selected once when its
master code is written and reused for subsequent prefix reads.

\subsection{Nested Storage and Reconstruction Compensation}

\begin{figure}[t]
    \centering
    \includegraphics[width=\linewidth]
        {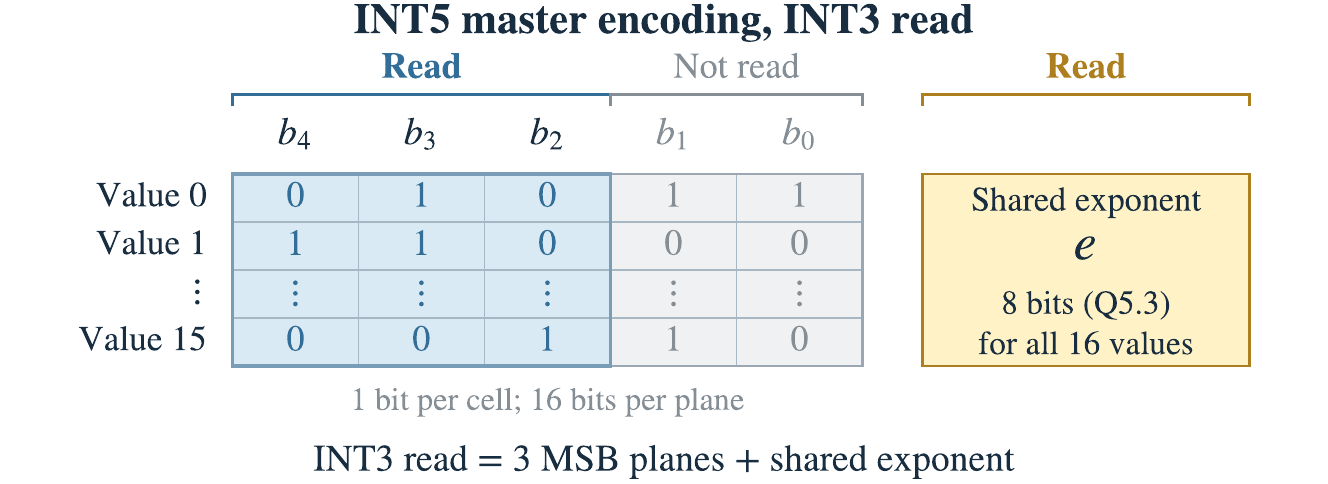}
    \caption{
Nested bit-plane storage and selective reading.
Sixteen INT5 values share one 8-bit Q5.3 exponent.
Rows represent values, and columns represent bit positions
from most to least significant.
An INT3 read fetches the three leading planes (blue)
and the exponent, leaving the two lower planes (gray)
unread.
Other widths select different numbers of leading planes
from the same encoding.
    }
    \label{fig:nested_storage}
\end{figure}

\phantomsection\label{flow:nested_reads}
Nested MXINT stores $w_{\max}$-bit master codes in
bit planes, with one 8-bit Q5.3 shared exponent $e$
per quantization group.
A width-$w$ read fetches only the top $w$ planes
and the exponent, supporting multiple precisions
from one weight or KV encoding.
Figure~\ref{fig:nested_storage} illustrates an INT3
read from INT5 master codes.

\phantomsection\label{flow:reconstruction}
With $d=w_{\max}-w$ bits omitted, the discarded suffix
ranges from $0$ to $2^d-1$.
Interpreting the fetched prefix $q_{\mathrm{hi}}$
as a signed integer, midpoint reconstruction gives
\begin{equation}
\widetilde q_w
=
2^d q_{\mathrm{hi}}+\frac{2^d-1}{2}.
\end{equation}
Hardware uses $q'_w=2\widetilde q_w$ and halves
the associated scale.
For a common read width,
\begin{equation}
\sum_j a_jq'_{w,j}
=
2^{d+1}\sum_j a_jq_{\mathrm{hi},j}
+
(2^d-1)\sum_j a_j,
\end{equation}
where $a_j$ is the other input's quantized integer.
Multipliers therefore process only narrow prefixes.
An accumulator at the first PE of each row computes
$\sum_j a_j$; shifts and subtraction generate a correction
shared across columns and added before scale alignment.

\phantomsection\label{flow:encoding}
For weights, we jointly optimize the shared exponent
$e$ and master codes $\boldsymbol q$ offline:
\begin{equation}
J(e,\boldsymbol q)
=
\sum_{w\in\mathcal W}
p_w\,
\delta_w(e,\boldsymbol q)^\top
H\,
\delta_w(e,\boldsymbol q).
\end{equation}
Here, $\delta_w$ is the weight reconstruction error
at width $w$, and $H=\mathbb E[xx^\top]$ is collected
from FP16 calibration inputs, both in the rotated basis.
Width-usage probabilities $p_w$ are measured through
mixed-precision calibration with initially quantized
weights, accounting for both input statistics and
read-width usage.

\subsection{Mixed-Precision Arrays and Reduction Trees}
\label{sec:mixed_precision_arrays}
Each compute core's $16\times2048$ flattened systolic
array comprises 128 $16\times16$ subarrays along
the reduction dimension, each with independently
configurable widths for its two inputs.
Each group of 16 input values shares a Q5.3 exponent.
Matching this group size to the subarray reduction
length keeps each PE's product scale fixed during
accumulation, avoiding local exponent alignment.
Different PEs may use different scales.

\phantomsection\label{flow:gating}
Bit-plane clock gating disables unused register bits,
while input isolation supplies valid signed inputs
at the physical datapath width.
Widths are configured before each subarray execution
and remain fixed throughout it, allowing consecutive
executions to change precision without width-switching
bubbles.

Corresponding PE outputs are aligned to a common scale
during cross-subarray reduction.
Shifts handle integer exponent differences, while
table-selected shift-and-add/subtract combinations
approximate the scaling factors for fractional differences.
The aligned integer results enter a 3:2 carry-save
compression tree, with carry propagation only at
the final stage.
Active-width gating and offset-binary representation
with an aggregate bias correction reduce switching
in unused upper bits and sign extensions.
The reduced results are dequantized to FP16 through
the VRAM input path.

\begin{table*}[t]
    \centering
    \caption{Perplexity and zero-shot accuracy across models.
    BDQ uses W4A4KV4 with FP16 Q/P;
    PLENA uses uniform INT4 and Ours a unified average
    4-bit budget across W/A/Q/P/K/V.
    Avg.\ is the six-task mean accuracy.
    Green bold marks the lowest perplexity and highest
    average accuracy among low-precision methods.}
    \label{tab:model_quality}

    \small
    \setlength{\tabcolsep}{1.9pt}

    \begin{tabular}{lcc*{9}{c}}
        \toprule
        \multirow{2}{*}{Model}
        & \multirow{2}{*}{Precision}
        & \multirow{2}{*}{Method}
        & \multicolumn{2}{c}{PPL ($\downarrow$)}
        & \multicolumn{7}{c}{Accuracy (\%, $\uparrow$)} \\
        \cmidrule(lr){4-5}
        \cmidrule(lr){6-12}
        & & &
        WikiText-2 & C4 &
        ARC-C & ARC-E & HellaSwag & LAMBADA &
        PIQA & WinoGrande & Avg. \\
        \midrule

        & FP16 & FP16
        & 6.14 & 9.45
        & 53.50 & 77.57 & 79.12 & 75.51
        & 80.74 & 72.93 & 73.23 \\

        \cmidrule(lr){2-12}

        & & BDQ (PQ16)
        & 6.84 & 10.97
        & 51.03 & 76.10 & 76.77 & 73.42
        & 78.57 & 72.88 & 71.46 \\

        & & PLENA (PQ4)
        & 6.95 & 11.04
        & 49.06 & 73.86 & 76.01 & 72.25
        & 77.53 & 69.06 & 69.63 \\

        \rowcolor{oursgreen}
        \cellcolor{white}\multirow{-4}{*}{Llama-3-8B}
        & \cellcolor{white}\multirow{-3}{*}{W4A4KV4}
        & \textbf{Ours (Avg4)}
        & \perfbest{6.58} & \perfbest{10.22}
        & 51.37 & 75.76 & 77.48 & 75.45
        & 79.38 & 73.95 & \perfbest{72.23} \\
        \midrule

        & FP16 & FP16
        & 2.86 & 7.17
        & 64.25 & 85.94 & 84.93 & 79.37
        & 84.44 & 80.74 & 79.95 \\

        \cmidrule(lr){2-12}

        & & BDQ (PQ16)
        & 3.52 & 7.63
        & 62.83 & 84.88 & 84.07 & 79.42
        & 84.01 & 79.56 & 79.13 \\

        & & PLENA (PQ4)
        & 4.12 & 8.47
        & 56.06 & 78.20 & 82.42 & 75.51
        & 81.34 & 72.69 & 74.37 \\

        \rowcolor{oursgreen}
        \cellcolor{white}\multirow{-4}{*}{Llama-3-70B}
        & \cellcolor{white}\multirow{-3}{*}{W4A4KV4}
        & \textbf{Ours (Avg4)}
        & \perfbest{3.34} & \perfbest{7.54}
        & 62.46 & 84.93 & 84.55 & 79.82
        & 84.00 & 79.64 & \perfbest{79.23} \\
        \midrule

        & FP16 & FP16
        & 8.58 & 30.36
        & 57.00 & 83.12 & 80.81 & 76.44
        & 82.64 & 72.69 & 75.45 \\

        \cmidrule(lr){2-12}

        & & PLENA (PQ4)
        & 9.92 & 31.82
        & 56.23 & 81.31 & 79.59 & 75.39
        & 81.77 & 69.69 & 74.00 \\

        \rowcolor{oursgreen}
        \cellcolor{white}\multirow{-3}{*}{Llama-4-Scout}
        & \cellcolor{white}\multirow{-2}{*}{W4A4KV4}
        & \textbf{Ours (Avg4)}
        & \perfbest{9.28} & \perfbest{30.84}
        & 55.46 & 81.52 & 80.31 & 76.62
        & 81.66 & 71.90 & \perfbest{74.58} \\
        \midrule

        & FP16 & FP16
        & 4.66 & 10.58
        & 64.08 & 79.04 & 87.70 & 72.00
        & 84.00 & 70.72 & 76.26 \\

        \cmidrule(lr){2-12}

        & & PLENA (PQ4)
        & 4.95 & 10.99
        & 61.01 & 76.64 & 86.64 & 69.22
        & 82.37 & 70.09 & 74.33 \\

        \rowcolor{oursgreen}
        \cellcolor{white}\multirow{-3}{*}{
    \shortstack[l]{Qwen3.8-Flash\\Next}
}
        & \cellcolor{white}\multirow{-2}{*}{W4A4KV4}
        & \textbf{Ours (Avg4)}
        & \perfbest{4.84} & \perfbest{10.74}
        & 60.67 & 77.57 & 87.32 & 72.17
        & 83.08 & 70.96 & \perfbest{75.30} \\
        \midrule

        & FP16 & FP16
        & 4.09 & 9.31
        & 60.24 & 81.06 & 86.09 & 74.52
        & 83.62 & 79.87 & 77.57 \\

        \cmidrule(lr){2-12}

        & & PLENA (PQ4)
        & 4.70 & 9.96
        & 59.81 & 81.14 & 84.91 & 74.03
        & 81.77 & 76.80 & 76.41 \\

        \rowcolor{oursgreen}
        \cellcolor{white}\multirow{-3}{*}{GLM-4.6V}
        & \cellcolor{white}\multirow{-2}{*}{W4A4KV4}
        & \textbf{Ours (Avg4)}
        & \perfbest{4.33} & \perfbest{9.50}
        & 57.94 & 79.80 & 85.72 & 74.67
        & 83.08 & 79.08 & \perfbest{76.71} \\

        \bottomrule
    \end{tabular}
\end{table*}

\begin{table}[t]
    \centering
\caption{Single- and multi-needle retrieval scores (\%)
on RULER, with FP16 prefill and method-specific decode
precision. Highlighted averages are best among
quantized methods.}
    \label{tab:long_context_retrieval}
    \small
    \setlength{\tabcolsep}{3pt}
    \begin{tabular}{@{}lcccccccc@{}}
        \toprule
        & \multicolumn{4}{c}{Single-needle}
        & \multicolumn{4}{c}{Multi-needle} \\
        \cmidrule(lr){2-5}\cmidrule(lr){6-9}
        Method
        & 32K & 64K & 128K & Avg.
        & 32K & 64K & 128K & Avg. \\
        \midrule

        \multicolumn{9}{l}{\textit{Llama-3.1-8B}} \\
        FP16
        & 100 & 99.3 & 97.3 & 98.9
        & 91.4 & 87.0 & 43.7 & 74.0 \\

        PLENA
        & 98.3 & 98.0 & 91.0 & 95.8
        & 84.2 & 64.6 & 24.5 & 57.8 \\

        \rowcolor{oursgreen}
        \textbf{Ours (avg4)}
        & 100 & 99.7 & 96.3 & \best{98.7}
        & 96.2 & 77.8 & 38.8 & \best{70.9} \\

        \midrule

        \multicolumn{9}{l}{\textit{Qwen3-8B}} \\
        FP16
        & 100 & 99.0 & 99.0 & 99.3
        & 99.4 & 97.8 & 93.3 & 96.8 \\

        PLENA
        & 98.7 & 94.3 & 95.3 & 96.1
        & 96.6 & 92.2 & 83.8 & 90.9 \\

        \rowcolor{oursgreen}
        \textbf{Ours (avg4)}
        & 100 & 100.0 & 98.0 & \best{99.3}
        & 99.3 & 97.6 & 93.5 & \best{96.8} \\

        \midrule

        \multicolumn{9}{l}{\textit{Qwen3-30B-A3B-Instruct}} \\
        FP16
        & 100 & 100 & 100 & 100
        & 100 & 100 & 99.8 & 99.9 \\

        PLENA
        & 100 & 100 & 100 & 100
        & 99.9 & 99.8 & 99.4 & 99.7 \\

        \rowcolor{oursgreen}
        \textbf{Ours (avg4)}
        & 100 & 100 & 100 & \best{100}
        & 100 & 100 & 100 & \best{100} \\

        \bottomrule
    \end{tabular}
\end{table}

\begin{table*}[t]
    \centering
    \caption{Decode performance at (input, output) token lengths.
    B200 results are measured on four GPUs;
    PLENA and Ours are modeled as 288-accelerator systems.
    All systems use their largest memory-feasible batches.
    Green bold marks the highest throughput and energy
    efficiency for each model and workload.}
    \label{tab:decode_performance}

    \small
    \setlength{\tabcolsep}{4pt}

    \begin{tabular}{ll rrr rrr rrr}
        \toprule
        \multirow{2}{*}{Model}
        & \multirow{2}{*}{System}
        & \multicolumn{3}{c}{(1.4K, 0.2K)}
        & \multicolumn{3}{c}{(114K, 5K)}
        & \multicolumn{3}{c}{(90K, 8K)} \\
        \cmidrule(lr){3-5}
        \cmidrule(lr){6-8}
        \cmidrule(lr){9-11}
        & &
        Tokens/s & Tokens/J & Batch
        & Tokens/s & Tokens/J & Batch
        & Tokens/s & Tokens/J & Batch \\
        \midrule

        \multirow{3}{*}{Llama-3.1-8B}
        & 4$\times$B200 NVFP4
        & 92,167 & 30.57 & 4,096
        & 3,913 & 1.01 & 147
        & 4,553 & 1.17 & 179 \\

        & PLENA
        & 153,060 & \perfbest{146.81} & 4,096
        & 7,030 & 4.66 & 147
        & 8,643 & 5.75 & 179 \\

        \rowcolor{oursgreen}
        & \textbf{Ours}
        & \perfbest{153,411} & 140.07 & 4,096
        & \perfbest{17,960} & \perfbest{12.16} & 119
        & \perfbest{21,619} & \perfbest{14.70} & 145 \\
        \midrule

        \multirow{3}{*}{Llama-3.1-70B}
        & 4$\times$B200 NVFP4
        & 34,136 & 9.03 & 3,919
        & 1,164 & 0.34 & 52
        & 1,679 & 0.44 & 63 \\

        & PLENA
        & \perfbest{40,027} & \perfbest{33.50} & 3,919
        & 2,514 & 1.71 & 52
        & 3,085 & 2.11 & 63 \\

        \rowcolor{oursgreen}
        & \textbf{Ours}
        & 39,919 & 30.48 & 3,170
        & \perfbest{5,175} & \perfbest{3.70} & 42
        & \perfbest{6,246} & \perfbest{4.42} & 51 \\
        \midrule

        \multirow{3}{*}{Llama-4-Scout}
        & 4$\times$B200 NVFP4
        & 38,322 & 13.90 & 4,096
        & 6,855 & 1.92 & 276
        & 8,048 & 2.22 & 323 \\

        & PLENA
        & 87,001 & \perfbest{78.23} & 4,096
        & 13,931 & 9.61 & 276
        & 16,118 & 11.20 & 323 \\

        \rowcolor{oursgreen}
        & \textbf{Ours}
        & \perfbest{87,173} & 74.38 & 4,096
        & \perfbest{24,358} & \perfbest{17.76} & 223
        & \perfbest{27,221} & \perfbest{19.64} & 261 \\
        \midrule

        \multirow{3}{*}{GLM-4.6V}
        & 4$\times$B200 NVFP4
        & 35,674 & 12.52 & 4,096
        & 1,666 & 0.65 & 90
        & 2,403 & 0.86 & 109 \\

        & PLENA
        & 95,579 & \perfbest{90.70} & 4,096
        & 4,331 & 2.97 & 90
        & 5,316 & 3.66 & 109 \\

        \rowcolor{oursgreen}
        & \textbf{Ours}
        & \perfbest{95,772} & 86.26 & 4,096
        & \perfbest{8,267} & \perfbest{6.02} & 73
        & \perfbest{10,131} & \perfbest{7.34} & 88 \\
        \midrule

        \multirow{3}{*}{Qwen3-235B-A22B}
        & 4$\times$B200 NVFP4
        & 25,850 & 7.52 & 4,096
        & 1,307 & 0.45 & 77
        & 1,857 & 0.58 & 94 \\

        & PLENA
        & 54,033 & \perfbest{56.90} & 4,096
        & 3,474 & 2.45 & 77
        & 4,239 & 3.01 & 94 \\

        \rowcolor{oursgreen}
        & \textbf{Ours}
        & \perfbest{54,374} & 54.27 & 4,096
        & \perfbest{4,759} & \perfbest{3.81} & 62
        & \perfbest{5,860} & \perfbest{4.67} & 76 \\

        \bottomrule
    \end{tabular}
\end{table*}

\section{Experiments}

\subsection{Experimental Setup}
\label{sec:experimental_setup}

\paragraph{a) Models and quality evaluation.}
We evaluate dense and MoE models from the
LLaMA~\cite{llama2,llama3}, Qwen, and GLM families.
Perplexity uses 2048-token windows on WikiText-2
test~\cite{wikitext2} and 256 C4 validation
windows~\cite{c4}.
Zero-shot evaluation uses
lm-evaluation-harness~\cite{lm_eval_harness} on
ARC-C/E~\cite{arc}, HellaSwag~\cite{hellaswag},
LAMBADA~\cite{lambada}, PIQA~\cite{piqa}, and
WinoGrande~\cite{winogrande}, reporting normalized
accuracy where available, otherwise accuracy,
and the six-task mean.
Baselines include BDQ~\cite{bdq}, PLENA~\cite{plena},
and FP16; precision settings appear in the result tables.
RULER~\cite{ruler} single- and multi-needle retrieval
uses 32K/64K/128K contexts and 300 samples per length,
following official generation and scoring protocols.
Coffee, FP16, and PLENA use FP16 prefill and
method-specific decode precision.
Calibration uses samples from C4, WikiText-2,
and Paul Graham essays.
Calibration and evaluation samples are disjoint.

\paragraph{b) Hardware implementation.}
Coffee and our PLENA reimplementation are synthesized
from SystemVerilog using Synopsys Design Compiler
and ASAP7~\cite{asap7} at 1\,GHz.
Each accelerator uses one $16\times2048$ core with
Q5.3 shared exponents; all PLENA results use this
reimplementation.
Logic power is estimated from VCD activity.
Energy includes compute/control logic, vector/scalar
units, precision selection, reconstruction, SRAM,
HBM at 3.5\,pJ/bit, static power, and communication.

\paragraph{c) System evaluation.}
We measure NVFP4 inference~\cite{nvfp4} and KV caching
on four B200 GPUs using vLLM~\cite{vllm} 0.31.0
with four-way tensor parallelism and prefix caching
disabled.
Matched one-token runs are subtracted to isolate
decode time and NVML board energy.
PLENA and Coffee are modeled as 288-accelerator systems,
approximately matching the estimated multiplier
count of four B200 GPUs.
Aggregate HBM bandwidth is matched at 32\,TB/s,
and available KV capacity is matched per model
to vLLM measurements.
PLENA and Coffee share workload mapping, communication
model, and simulator.
All systems store one copy of the weights and run
the same PG-19 decode workloads at their maximum
memory-feasible batch sizes.
We report throughput, energy efficiency, and HBM traffic.

\subsection{Software--Hardware Co-Design}
\label{sec:precision_area_codesign}
\begin{figure}[t]
    \centering
    \includegraphics[width=\linewidth]
        {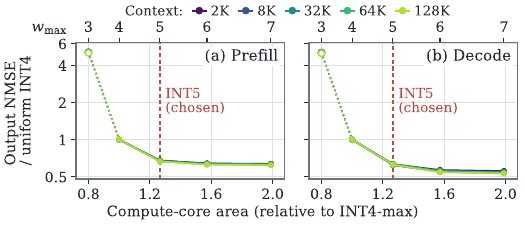}
    \caption{Precision--area trade-offs on Llama-3.1-8B.
    NMSE and compute-core area (SAs and CSA reduction tree)
    are normalized to uniform INT4 and the INT4-capable
    design, respectively.
    Top axes indicate $w_{\max}$; hollow markers denote
    INT3 configurations with unused budget, and dashed
    vertical lines mark the selected INT5 design.}
    \label{fig:precision_area}
\end{figure}
We evaluate maximum widths from INT3 to INT7, with
INT2 as the minimum width and an average 4-bit budget.
Figure~\ref{fig:precision_area} reports normalized
output NMSE averaged across nine linear and attention
GEMMs from all 32 layers of Llama-3.1-8B, at five
context lengths from 2K to 128K.
We select INT5 as the maximum supported width.
Averaged across the five contexts, online allocation
reduces output NMSE by 32.9\% for prefill and 37.2\%
for decode, with 26.6\% more compute-core area
than the INT4-capable design.
INT5 captures 88.5\% and 81.0\% of the respective
NMSE reductions achieved with INT7, whose area
is 98.9\% larger than the INT4-capable design.

\subsection{Model Quality}
\label{sec:model_quality}

Coffee achieves lower perplexity and higher mean
zero-shot accuracy across available quantized-baseline
comparisons in Table~\ref{tab:model_quality}.
On Qwen3.8-Flash, mean accuracy is 75.30\%,
0.96 percentage points below FP16.
Single- and multi-needle retrieval matches or improves
on PLENA across all three evaluated models
(Table~\ref{tab:long_context_retrieval}).

\subsection{System Performance and Energy Efficiency}
\label{sec:system_performance}

At short context, INT5 weight/KV storage and metadata
increase memory usage and can limit batch sizes,
while KV-read savings remain modest.
Throughput is comparable to PLENA, but energy
efficiency is approximately 5--9\% lower.
At long context, greater KV-read savings provide
larger benefits, with V skipping and mixed-precision
KV reads contributing approximately 55\% and 45\%
of the total KV traffic reduction, respectively.
Coffee achieves up to $2.55\times$ PLENA's throughput
and $2.61\times$ its energy efficiency, and up to
$4.96\times$ and $12.56\times$, respectively,
compared with four B200 GPUs
(Table~\ref{tab:decode_performance}).

\subsection{Online Precision-Decision Overhead}
\label{sec:decision_overhead}

Table~\ref{tab:decision_overhead} shows low area and
energy overheads: the estimator and PSU occupy 1.97\%
of core area.
Online decisions change prefill time by less than
0.01\% and increase the first decode step's time
by only 0.31--0.64\%.

\begin{table}[t]
    \centering
    \caption{Online precision-decision overhead on
    Llama-3.1-8B.
    (a) Area and energy as percentages of core area
    and the corresponding phase's total energy.
    (b) Execution-time overhead; decode measures
    the first step after prefill.}
    \label{tab:decision_overhead}
    \label{tab:decision_stalls}
    \footnotesize
    \setlength{\tabcolsep}{3pt}

    \begin{tabular*}{\linewidth}{@{\extracolsep{\fill}}lccc@{}}
        \toprule
        \multicolumn{4}{@{}l}{\textbf{(a) Area and energy}} \\
        \midrule
        Component
        & \shortstack{Area\\(\% of core)}
        & \shortstack{Prefill\\energy (\%)}
        & \shortstack{Decode\\energy (\%)} \\
        \midrule
        Error estimator
        & 0.33 & 0.011--0.116 & 0.012--0.019 \\
        PSU
        & 1.64 & 0.482--0.491 & 0.123--0.231 \\
        \midrule
        Total logic
        & 1.97 & 0.49--0.61 & 0.135--0.25 \\
        \midrule
        HBM: error statistics
        & -- & 0.27--0.48 & 1.75--3.09 \\
        HBM: attention mass
        & -- & -- & 0.11--0.19 \\
        \bottomrule
    \end{tabular*}

    \par\medskip

    \begin{tabular*}{\linewidth}{@{\extracolsep{\fill}}lcccc@{}}
        \toprule
        \multicolumn{5}{@{}l}{
            \textbf{(b) Execution-time overhead (\%)}
        } \\
        \midrule
        Phase & 8K & 32K & 64K & 90K \\
        \midrule
        Prefill
        & $<0.01$ & $<0.01$ & $<0.01$ & $<0.01$ \\
        Decode
        & 0.31 & 0.54 & 0.64 & 0.62 \\
        \bottomrule
    \end{tabular*}
\end{table}

\subsection{Component Evaluation}
\label{sec:component_ablation}

\begin{figure}[t]
    \centering
    \includegraphics[width=0.89\linewidth]
        {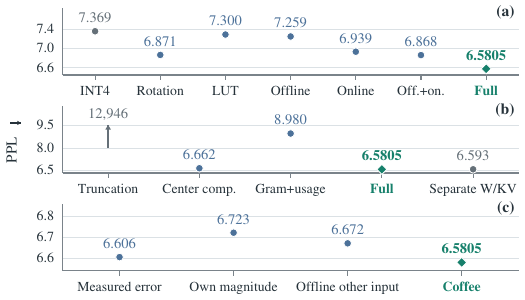}
    \caption{WikiText-2 perplexity on Llama-3-8B.
    (a) Component ablation under an average 4-bit budget.
    LUT denotes adaptive scale selection; Full combines
    all components.
    (b) Nested encoding ablation with a fixed width schedule.
    The arrow marks an off-scale value.
    (c) Width-selection criteria with other settings fixed.
    Measured error uses directly computed mean-square
    magnitudes and reconstruction MSEs in the same proxy.
    Own magnitude uses only each input's magnitude,
    preserving Coffee's proportions at each width.
    Offline other input replaces the other input's current
    mean-square magnitude with its per-tile calibration average.
    Coffee uses estimates from both current inputs.}
    \label{fig:component_ablation}
\end{figure}

\paragraph{Model-quality ablation.}
Under an average 4-bit budget, online error-guided
allocation within offline-assigned budgets reduces
WikiText-2 perplexity on Llama-3-8B from 7.369 to 6.868,
comparable to rotation alone (6.871).
Adding rotation and adaptive scale selection further
lowers perplexity to 6.5805
(Figure~\ref{fig:component_ablation}(a)).
Using estimates from both current inputs yields lower
perplexity than using each input's magnitude alone
or offline averages for the other input, while achieving
perplexity comparable to using directly computed
magnitudes and reconstruction MSEs
(Figure~\ref{fig:component_ablation}(c)).

\paragraph{Nested storage.}
Center compensation reduces perplexity from 12,946
with direct truncation to 6.662; joint Gram- and
usage-aware optimization further lowers it to 6.5805,
close to 6.593 with independent W/KV quantization
(Figure~\ref{fig:component_ablation}(b)).

We also evaluate the read efficiency of nested encodings
at 128K context.
Without skipping, non-nested INT4 and nested reads
achieve 90\% and approximately 80\% of peak bandwidth,
respectively.
Despite the lower bandwidth utilization, combining
nested reads with V skipping substantially reduces
KV traffic, shortening KV read time by 57\%
relative to non-nested INT4.

\section{Conclusion}
\label{sec:conclusion}

Coffee uses both current inputs to allocate precision
within offline-assigned budgets.
Dedicated circuitry, nested reads, and gating support
efficient execution, improving model quality and
long-context inference efficiency under an average
4-bit budget.

\clearpage

\bibliography{references}
\bibliographystyle{mlsys2026}
\clearpage
\section*{AI Usage Disclosure}

We used Claude Code and OpenAI Codex to assist with
literature search and comparison, experimental design
discussions, code development, and language polishing
of the manuscript.
The authors retain full responsibility for the
methodology, implementation, experimental results,
references, and final manuscript.

\end{document}